\documentclass[journal=jpclcd,manuscript=letter,superscriptaddress]{achemso}

\usepackage{amsmath}
\usepackage{amsfonts}
\usepackage{amssymb}
\usepackage{graphicx}
\usepackage{epstopdf}
\usepackage{pstool}
\usepackage{subcaption}
\usepackage{siunitx}
\usepackage{booktabs, tabu} % tables
\usepackage{multirow}
\usepackage{mathrsfs}
\usepackage{cases}
\usepackage{color}
\usepackage{txfonts}
\author{Chuan-Cun Shu} \email{c.shu@unsw.edu.au}
\affiliation{School of Engineering and Information Technology,
University of New South Wales, Canberra, ACT 2600, Australia}
\author{Kai-Jun Yuan}
\affiliation{Laboratoire de Chimie Th\'{e}orique, Facult\'{e} des Sciences, Universit\'{e} de Sherbrooke
Sherbrooke (Qu\'{e}bec)  J1K 2R1,  Canada}
\alsoaffiliation{School of Engineering and Information Technology, University of New South Wales, Canberra, Australian Capital Territory 2600, Australia}
\date{\today}
\author{Daoyi Dong}
\affiliation{School of Engineering and Information Technology, University of New South Wales, Canberra, Australian Capital Territory 2600, Australia}
\alsoaffiliation{Department of Chemistry, Princeton University, Princeton, New Jersey 08544, USA}
\author{Ian R. Petersen}
\affiliation{School of Engineering and Information Technology,
University of New South Wales, Canberra, ACT 2600, Australia}
\author{Andre D.  Bandrauk}\email{andre.bandrauk@usherbrooke.ca}
\affiliation{Laboratoire de Chimie Th\'{e}orique, Facult\'{e} des Sciences, Universit\'{e} de Sherbrooke
Sherbrooke (Qu\'{e}bec)  J1K 2R1,  Canada}
\title[An \textsf{achemso} demo]
  {Identifying  Strong-Field Effects in Indirect Photofragmentation Reactions}

\begin{document}

%%%%%%%%%%%%%%%%%%%%%%%%%%%%%%%%%%%%%%%%%%%%%%%%%%%%%%%%%%%%%%%%%%%%%
%% The "tocentry" environment can be used to create an entry for the
%% graphical table of contents. It is given here as some journals
%% require that it is printed as part of the abstract page. It will
%% be automatically moved as appropriate.
%%%%%%%%%%%%%%%%%%%%%%%%%%%%%%%%%%%%%%%%%%%%%%%%%%%%%%%%%%%%%%%%%%%%%

 \begin{abstract}
Exploring molecular breakup processes induced by light-matter interactions has both fundamental and practical implications. However, it remains a challenge to elucidate  the underlying reaction mechanism in the strong field regime, where the potentials of the reactant are modified  dramatically.   Here,  we perform a theoretical analysis combined with a time-dependent wavepacket calculation to show how a strong ultrafast laser field affects the photofragment products. As an example, we examine the photochemical reaction of breaking up the molecule NaI into  the neutral atoms Na and I, which due to inherent nonadiabatic couplings is indirectly formed in a stepwise fashion via the  reaction intermediate  NaI$^{\ast}$. By analyzing the angular dependencies of fragment distributions,  we are able to identify the reaction intermediate  NaI$^{\ast}$  from the  weak to the strong field-induced nonadiabatic regimes. Furthermore, the energy levels of NaI$^{\ast}$ can be extracted from the  quantum interference patterns of the transient photofragment momentum distribution.
\end{abstract}

%%%%%%%%%%%%%%%%%%%%%%%%%%%%%%%%%%%%%%%%%%%%%%%%%%%%%%%%%%%%%%%%%%%%%
\maketitle
Since the scientist Ahmed Zewail (1946-2016) was awarded the 1999 Nobel Prize in Chemistry for his work on femtochemistry, controlling chemical reactions by employing light as a ``photonic reagent"  has become  a long-standing target going beyond traditional approaches using heat or catalysts \cite{jpca:104:5660,nature:348:225,science:288:824,nature:515:45,nature:515:100,pnas:112:15613}.  One type of photochemical reaction  is the dissociation of a molecule into fragments \cite{PD-Schinke},  which for example plays an important role in energy transfer processes during the light reaction of photosynthesis \cite{nature:446:782}. There has been considerable  theoretical and experimental interest in the study of such processes in both diatomic and  polyatomic molecular systems \cite{science:314:278,naturechem:6:785,pccp:4:5554,jpcl:5:3854,jpcl:6:824,jpca:116:11434,jcp:126:134306,jpcl:2:1715,jcp:136:174303}. Photodissociation dynamics can be broadly classified as direct and indirect processes \cite{PD-Schinke}, which can be identified by using a potential energy surface concept.
For a direct fragmentation reaction, an excited wavepacket evolves following  a purely repulsive potential, from which the molecule can fly apart immediately on  an ultrafast time-scale smaller than a typical internal vibrational period. For an indirect fragmentation reaction,  the potential of the excited state involved is not purely repulsive, where the  excited wavepacket  may survive for a sufficiently long time from several to thousands of internal vibrational periods.  \\ \indent
 It is very difficult to gain an insight into  photochemical reactions in the presence of strong fields, where the molecular Hamiltonian associated with electronic, vibrational and rotational energy levels is significantly modified by the applied fields. By considering field-dressed Born-Oppenheimer potentials, the concept of  light-induced potential (LIP)  \cite{jcp:68:3040,jcp:74:1110,jcp:97:12620,jcp:130:124320} is  often used to interpret various strong-field-induced phenomena including bond softening \cite{prl:64:1883}, bond hardening  \cite{prl:83:3625} and  above-threshold dissociation \cite{prl:64:515,prl:98:163001}. An intriguing approach  regarding LIPs investigates  the creation of light-induced conical intersection (LICI) in diatomic molecules \cite{jpcl:6:348,jpb:41:221001,jpb:44:045603}, which cannot form a natural conical intersection (CI) under field-free conditions. Considerable theoretical approaches have examined the impact of LICI on the direct fragmentation reaction \cite{prl:106:123001,jpca:116:2636,jpca:117:2636,jpca:118:11908,fd:va}, and has demonstrated the existence of LICI in diatomic molecules \cite{prl:116:143004}. However, these strong-field-induced phenomena  are largely unexplored in the case of indirect fragmentation reactions. In this Letter,  we show how a strong ultrafast laser pulse affects indirect photofragment distributions of  diatomic molecules in the photochemical reaction AB $\xrightarrow{laser}$ (AB)$^{\ast}$ $\rightarrow$ A+B, which in turn provides a new approach to identify quantum energy level structures of the reaction intermediate  AB$^{\ast}$ from photochemical reaction products. \\ \indent
As a prototype system, we focus on the indirect photodissociation reaction of the molecule NaI, which consists of the ground ionic state of $\rm{Na^{+}I^{-}}$ and the lowest covalent excited state NaI.  The corresponding diabatic  potential curves, ground $V_g(R)$ and excited $V_e(R)$,  interact with each other nonadiabatically  around their crossing at an internuclear distance of $R_c\approx6.93$ {\AA} \cite{jcp:91:7415}, where an avoided covalent-ionic curve crossing via a nondiabatic coupling $V_{c}(R)$ is formed.  We consider the molecule to be excited by a linearly polarized  ultrafast laser pulse, whose electric field vector  is described by  $\mathcal{\mathbf{E}}(t)=\hat{\mathbf{e}}_z\mathcal{E}_0f(t)\cos\omega_0t$ with the carrier frequency $\omega_0$, peak amplitude  $\mathcal{E}_0$, pulse envelope $f(t)$  and polarization direction $\hat{\mathbf{e}}_z$.  The molecular Hamiltonian $\hat{H}(t)$ using the electric-dipole approximation can be given  by
\begin{eqnarray}
\hat{H}(t)&=&   \left(\begin{array}{cc}
 -\frac{1}{2\mu}\frac{\partial^2}{\partial R^2}+\frac{\hat{J}^2}{2\mu R^2} & 0 \\
 0 & -\frac{1}{2\mu}\frac{\partial^2}{\partial R^2}+\frac{\hat{J}^2}{2\mu R^2} \\
 \end{array}
 \right)\nonumber\\ \
                                                                                                     && +\left(
                                                                                                        \begin{array}{cc}
                                                                                                         V_g(R) & -d(R)\cos\theta\mathcal{E}_0f(t)\cos\omega_0t +V_{c}(R) \\
                                                                                                          -d(R)\cos\theta\mathcal{E}_0f(t)\cos\omega_0t +V_{c}(R) & V_e(R) \\
                                                                                                        \end{array}
                                                                                                      \right), \label{Ht}
\end{eqnarray}
where $\hat{J}^2$ is the angular momentum operator of the nuclear rotation,  $\theta$  the angle between the molecular axis and the laser field polarization direction $\hat{\mathbf{e}}_z$, and $d(R)$ the transition dipole moment between the two diabatic electronic states. After diagonalizing the diabatic potential energy matrix in Eq. (\ref{Ht}), two time-dependent LIPs,  $V_g^{LIP}(R, \theta, t)$ and $V_e^{LIP}(R,\theta, t)$  can be written in the adiabatic representation as
\begin{subequations} \label{LIP}
\begin{equation}\label{LIPg}
 V_g^{LIP}(R,\theta,t)=\frac{1}{2} \left[V_e(R)-\hbar\omega_0+ V_g(R)-\sqrt{\Delta^2+4C^2(t)}\right],
 \end{equation}
 and
 \begin{equation}\label{LIPe}
   V_e^{LIP}(R,\theta,t)=\frac{1}{2} \left[V_e(R)-\hbar\omega_0+ V_g(R)+\sqrt{\Delta^2+4C^2(t)}\right],
 \end{equation}
\end{subequations}
where the detuning $\Delta$ is given by $\Delta=V_e(R)-\hbar\omega_0- V_g(R)$, and the total nondiabatic coupling $C(t)$ is defined as  $C(t)=-\frac{1}{2}d(R)\mathcal{E}_0f(t)\cos\theta +V_{c}(R)$. A detailed description of diabatic and adiabatic representations can be found in Ref \cite{mp:19:95}. The two LIPs are degenerate,  whenever $\Delta=0$ and $C(t)=0$  are fulfilled simultaneously. As the nondiabatic coupling $V_c(R)$ for the molecule NaI is negligible in the Franck-Condon region, the  condition  $C(t)=0$  always holds at  $\theta=\pi/2$ (i.e., $\cos\theta=0$). As a result, the LICI can be formed in the covalent  molecule NaI, and its position is determined  by the  carrier frequency $\omega_0$.\\ \indent
\begin{figure}[!t]\centering
\resizebox{0.8\textwidth}{!}{%
  \includegraphics{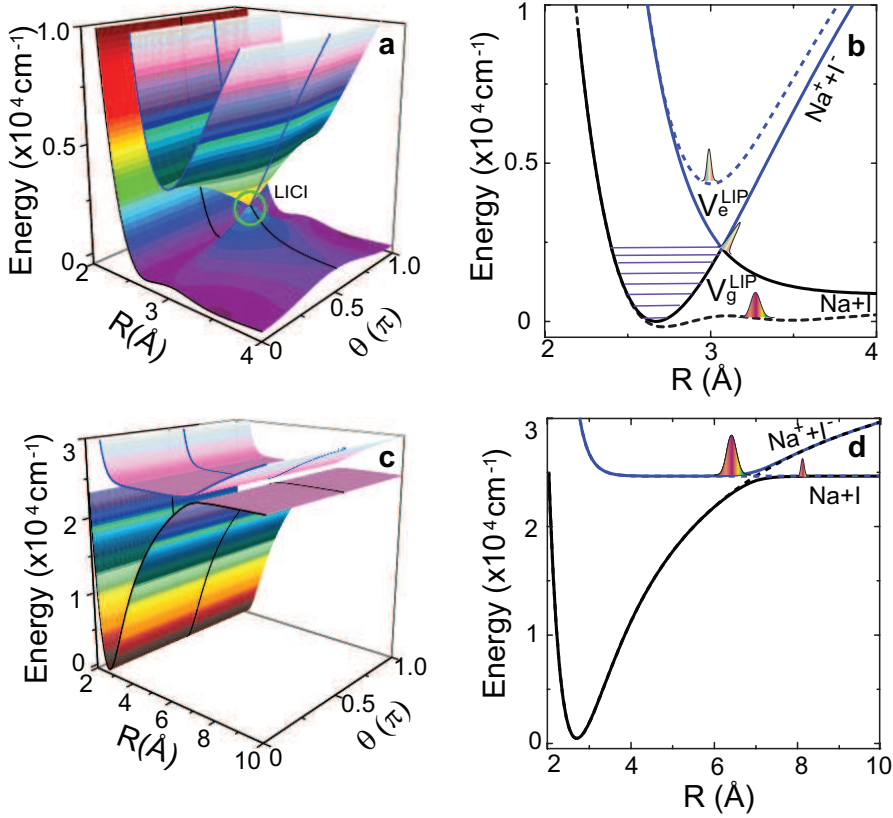}
} \caption{Schematic illustration of the photodissocation reaction of NaI. \textbf{a} The light-induced potentials (LIPs)  and \textbf{c} the field-free potential energy surfaces  of the molecule NaI as a function of the internuclear distance $R$ and the angle $\theta$ between the molecular axis and the laser polarization direction.  \textbf{b} A cut through the LIPs at $\theta=0$ (dashed lines ) and $\pi/2$ (solid lines), where $V_g^{LIP}$ and $V_e^{LIP}$ are coloured in dark and blue, respectively. \textbf{d} The diabatic potentials for the ground ionic state (black dashed line) and the lowest covalent excited state (the blue dashed line), and the adiabatic potentials for the ground electronic state (black solid line) and the excited electronic state (blue solid line). The green circle in \textbf{a} highlights the light-induced conical intersection (LICI) region. The wave packet trapped in the upper  adiabatic potential corresponds to the reaction intermediate  NaI$^{\ast}$.}\label{fig1}
\end{figure}
  Figure \ref{fig1}a shows $V_g^{LIP}(R, \theta)$ and $V_e^{LIP}(R,\theta)$ at $\hbar\omega_0=24390$ cm$^{-1}$ (410 nm). The gap between the two LIPs at a given internuclear distance $R$ is maximal at the angles $\theta=0$ and $\pi$, whereas it vanishes  at the angle  $\theta=\pi/2$ and the internuclear distance  $R=R_{CI}\approx3.1$ {\AA}.    To explore how the two different LIPs as well as the LICIs play roles,  as sketched in Figure \ref{fig1} b,  we  prepare the molecular system initially in a high vibrational level of the ground electronic state, so that  the excited wavepackets  evolve  simultaneously on both $V_g^{LIP}$ and   $V_e^{LIP}$. The lower LIP $V_g^{LIP}$ correlates $V_g$  to  $V_e$ at large $R$ ($>R_{CI})$, whereas the upper LIP $V_e^{LIP}$ connects $V_g$ to $V_e$ at small $R$ ($<R_{CI}$).  Since there is no internal barrier,  the field-induced bound states are created in $V_e^{LIP}$, resulting in a vibrational trapping, i.e., chemical bond hardening \cite{prl:83:3625}. In the vicinity of the LICIs, the gap between the two LIPs is small, giving rise to a nonadiabatic transition to the resonance states of $V_e$, which is similar to the weak-field one-photon resonance excitation.\\ \indent
We first analytically examine how the excited wavepacket evolves after these LIPs as well as LICIs disappear.  Under the field-free condition the  inherent nondibatic coupling $V_c(R)$  results in two adiabatic curves $V_g^{ad}(R)$ and $V_e^{ad}(R)$ with a trapping well in the upper excited state, as shown in Figures \ref{fig1}c and d. Different from the direct fragmentation reaction, the excited nuclear wavepacket, i.e., the reaction intermediate  NaI$^{\ast}$,   performs oscillating motions in the excited adiabatic electronic state $V_e^{ad}$, resulting in indirect dissociation along the Na+I channel in a stepwise fashion  \cite{nature:348:225,jcp:91:7415}.  The coherent superposition of quasibound states NaI$^{\ast}$  can be expanded as
 \begin{equation}\label{psiead}
  |\Psi_e^{ad}(t)\rangle=\sum_{\nu,J}c_{\nu J}|E_{\nu}\rangle e^{-iE_{\nu}t/\hbar}|E_{JM}\rangle e^{-iE_{JM}t/\hbar},
   \end{equation}
where $|E_{\nu}\rangle$ is  the vibrational  eigenstate with the vibrational quantum number $\nu$,  $|E_{JM}\rangle$ is the rotational eigenstate with the rotational quantum numbers $J$ and $M$,  $c_{\nu J}$ denotes the corresponding expansion coefficient of the rovibrational state $|E_\nu E_{JM}\rangle$.  For linearly polarized light case, the quantum number $M$ is conserved, and therefore the rotational state $|E_{JM}\rangle$ of the molecule can be described by the associated-Legendre polynomial $P_J^M(\cos\theta)$. When such a wavepacket approaches the nonadiabatic coupling region $R\approx R_c$, the quasibound states will  decay to the continuum via the nondiabatic coupling $V_c$. The ``leaked out" wave packet in the continuum states can be described by \cite{jcp:134:164308,jcp:132:234311}
$|\Psi_e(t)\rangle=-\frac{i}{\hbar}\int_{0}^{t}dt'\hat{U}_0(t,t')V_c|\Psi_e^{ad}(t)\rangle,$
 where $\hat{U}_0(t,t')$ is the field-free evolution operator of the system.\\ \indent
  The probability of observing fragments in an eigenstate $|E, p\rangle$ along the direction $\theta$ at the time $t$ can be given by
  \begin{eqnarray}\label{PE1}
P(E, \theta, t)&\equiv&\left|\langle E, p|\Psi_e(\theta,t)\rangle\right|^2 \nonumber\\ \
&\propto&\left|\sum_{\nu J}c_{\nu J}V_{{\nu JM}E}\frac{e^{-(E_{\nu}+E_{JM})t/\hbar}-1}{(E_{\nu}+E_{JM}-E)/\hbar}P_{J}^M(\cos\theta)\right|^2
   \end{eqnarray}
 where $V_{{\nu JM}E}=\langle E, p|V_c|E_{\nu}E_{JM}\rangle$ denotes the coupling matrix elements between the resonant state $|E_{\nu}E_{JM}\rangle$ and the continuum state $|E, p\rangle$, with the total energy of the fragments $E=p^2/2\mu+D_0$ and the dissociation energy $D_0$ of  Na+I. A further analysis of Eq. (\ref{PE1}) predicts that the  energy distribution consists of multiple peaks centered at $E_{\nu}+E_{JM}$ weighted by the expansion coefficient $c_{\nu J}$ and the coupling matrix element $V_{{\nu JM}E}$ at the corresponding energy, and such a time-dependent distribution  exhibits low energy resolution at short times \cite{jcp:134:164308,jcp:132:234311}. The components of the excited resonance states $|E_{\nu}E_{JM}\rangle$ of NaI$^{\ast}$  in the weak light-molecule interaction regime can be approximated  via first-order perturbation theory by
 \begin{equation}\label{cn}
c_{\nu J}=\langle E_{JM}E_{\nu}|d\cos\theta|\epsilon_0\rangle\int_{-\infty}^{\infty}dt\mathcal{E}_0f(t)\cos\omega_0te^{i\omega_{E_{\nu J0}}t},
   \end{equation}
 where $\omega_{E_{\nu J0}}=(E_{\nu}+E_{JM}-\epsilon_0)/\hbar$, and $|\epsilon_0\rangle$ is the initial rovibrational state of the system with the eigenenergy $\epsilon_0$, and therefore are determined by a product of the Franck-Condon factor and the  laser pulse frequency distribution at the transition frequency. In the presence of strong light-molecule interactions, the excited quasibound states $|E_{\nu}E_{JM}\rangle$ will be modified going beyond first-order perturbation theory, but as seen from Eq. (\ref{PE1}) the expansion coefficients $c_{\nu J}$ can be extracted from indirect photofragment distributions.\\ \indent
In the following we  perform a time-dependent quantum wavepacket calculation to examine  how such an excited wavepacket, i.e., the reaction intermediate  NaI$^{\ast}$,  is unfolded in real time.
The propagation of the wavepackets $\Psi_g(R, \theta, t)$ and $\Psi_e(R, \theta,t)$ with the two diabatic potentials $V_g(R)$ and $V_e (R)$ is  obtained by solving the time-dependent Schr\"{o}dinger equation (TDSE) with  the Hamiltonian in Eq. (\ref{Ht}). The linearly polarized laser field $\mathcal{\mathbf{E}}(t)$ is taken to be  an experimentally accessible Gaussian  transform limited pulse  centered at $t=0$ with the full-width at half-maximum (FWHM) of $30$ fs. The center wavelength of the laser pulse is fixed at 410 nm, and throughout the calculations the initial nuclear wave function was assumed to be in a single eigenstate of the ground electronic state with the vibrational quantum number of $\nu=7$ and the rotational quantum number of $J=0$.  The wave functions, potentials, and coupling element are represented on an equally spaced grid of 8192 points with $1.5\leq R \leq93$ {\AA}. An absorbing potential is added for $R \geq92$  {\AA} to avoid unphysical reflections into the inner region. The data for the potential energy surfaces $V_g(R)$ and $V_e(R)$, the transition dipole moments $d(R$ and the nondiabatic coupling $V_c(R)$ can be found in literature \cite{jcp:91:7415}. Since the used laser frequency is far from resonances with vibrational and rotational transitions within the same electronic state, the transitions induced by the permanent dipole moments are negligible. The further details for numerically solving TDSE   can be found in Refs \cite{jcp:134:164308,pra:79:023418}.  \\ \indent
The time-dependent momentum distribution of the photofragments NaI$^{\ast}\rightarrow$ Na+I is computed by Fourier transforming the wavefunction  $\Psi(R, \theta, t)$ from  position space to momentum space as
\begin{equation}\label{MD}
  \Phi(p, \theta, t)=\frac{1}{\sqrt{2\pi\hbar}}\int_{-\infty}^{\infty}\exp\left(-\frac{ipR}{\hbar}\right)g(R)\Psi_e(R, \theta, t)dR,
   \end{equation}
where $g(R)$ is a filter function that will effectively suppress the part of the wavepacket in  the avoided crossing region and becomes a constant of 1 in the asymptotic dissociation region. In our simulations, we take $g(R)=1-\exp[-a(R-R_d)]$ for $R>R_d$ and $g(R)=0$ for $R<R_d$ with $a=0.4$ {\AA}$^{-1}$ and $R_d=10$ {\AA}.
Such  time-dependent momentum distributions can be measured practically in experiment based on the technology of transient momentum imaging of photofragments.\\ \indent
\begin{figure*}[!t]\centering
\resizebox{0.8\textwidth}{!}{%
  \includegraphics{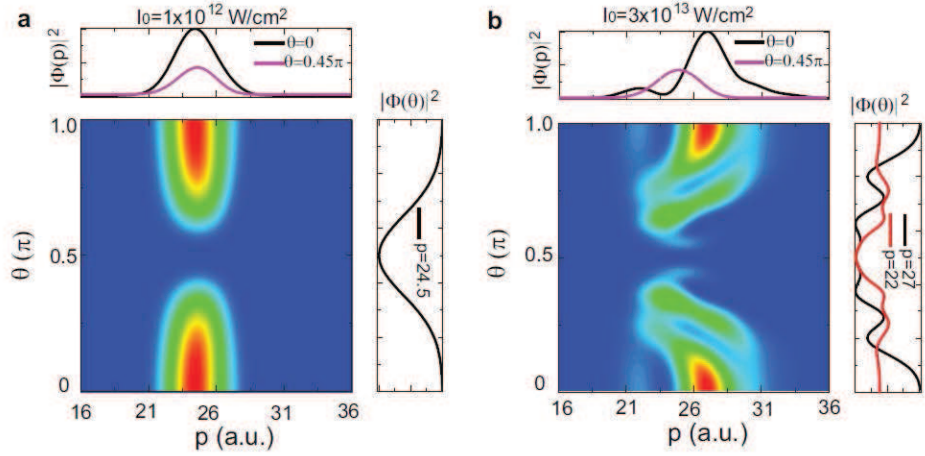}
} \caption{Probability distributions for relative momentum of Na+I. The results, i.e., $|\Phi(p, \theta, t)|^2$, are observed at  $t=0.8$ ps, corresponding to one outgoing fragment, with  \textbf{a} the weak-field case at $I_0=1.0\times10^{12}$ W/cm$^2$, and \textbf{b}  the strong field case at $I_0=3.0\times10^{13}$ W/cm$^2$.}\label{fig2}
\end{figure*}
\begin{figure}[!t]\centering
\resizebox{0.8\textwidth}{!}{%
  \includegraphics{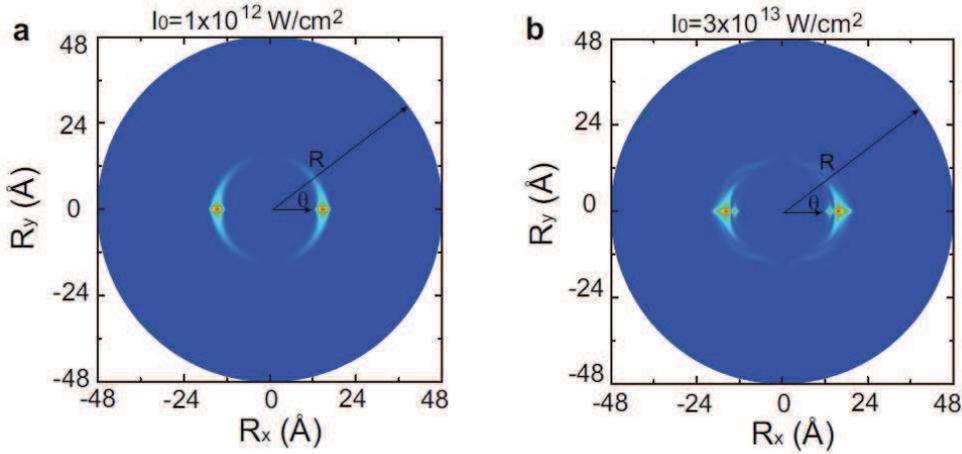}
} \caption{Probability distributions for the wavepackets of photofragments. The results are observed in the asymptotic dissociation region $|g(R)\Psi_e(R, \theta, t)|^2$ at $t=0.8$ ps, with  \textbf{a} the weak-field case at $I_0=1.0\times10^{12}$ W/cm$^2$, and \textbf{b}  the strong field case at $I_0=3.0\times10^{13}$ W/cm$^2$.}\label{fig3}
\end{figure}
Figure \ref{fig2} shows the angular distributions of the fragment momentum at  $t=0.8$ ps, corresponding to one outgoing fragment. For comparison, the simulations are accomplished by using a weak laser pulse at the peak intensity of  $I_0=1.0\times10^{12}$ W/cm$^2$ and  a strong one at $I_0=3.0\times10^{13}$ W/cm$^2$, respectively. In the weak field case,  the initial rovirational state $|\nu=7, J=0\rangle$ via one-photon transitions is coupled to the resonant state $|E_\nu\rangle$ with $\Delta J=\pm1$, which as predicted in Eq. (\ref{PE1}) produces a  $\cos^2\theta$ (i.e., $|P_1^0(\cos\theta)|^2$) distribution. As seen from Figure \ref{fig2}a, the fragment momentum  exhibits a Gaussian-like distribution with the maximum value around $p=24.5$ a. u. (i.e., $\sqrt{2\mu(\epsilon_0+\hbar\omega_0-D_0)}$ ), in good agreement with the theoretical analysis in Eqs. (\ref{PE1}) and (\ref{cn}). The photofragment distributions are obviously changed in the strong field case as compared with those  in  the weak-field case, see Figure \ref{fig2}b.  Two separated peaks around $p=22$ and 27 a.u. are produced in the strong laser-induced  nonadibatic coupling regions ($\theta\rightarrow 0,$ and $\pi$), whereas a Gaussian-like envelop with the maximum around $p=24.5$ a.u. appears again in the weak coupling region ($\theta\rightarrow \pi/2$).\\ \indent
The photofragment distribution as indicated in Eq. (\ref{PE1}) directly reflects the quasibound states of the excited electronic states. In Figure \ref{fig2}b the two shifted and separated peaks thus imply that the quasibound states below and above the resonant states are excited simultaneously. In the LIP representation, the blue-shifted peaks around $p=27$ a.u. correspond to the product from the lower LIP $V_g^{LIP}$ passage, which brings the system from the initial state to the higher quasibound states of $V_e$.
 The red-shifted peaks around $p=22$ a.u. comes from the upper LIP $V_e^{LIP}$, which moves the system from the initial state to those quasibound states below resonant states. The shifted value of the momentum at a given direction $\theta$ can be predicated by  Eq. (\ref{LIP}) with $\Delta p=\pm\sqrt{\mu d(R)\mathcal{E}_0\cos\theta}$ at $R=R_{CI}$.  Figure \ref{fig3} plots the probability density distribution  of the dissociating fragments $|g(R)\Psi_e(R, \theta, t)|^2$ on a ``disk" for both the weak and strong field cases. Comparing to the weak field case with a single-``crescent-like" distribution (Figure \ref{fig3}a), a double-``crescent-like" distribution around $\theta=0$ and $\pi$ (Figure \ref{fig3}b) is induced in the strong field case, indicating that two  different manifolds of quasibound states are included. In addition, both the momentum and probability density distributions consistently establishes that the fragmentats from the higher quasibound states dominates. This can be attributed to two main reasons: one is that more initial wavepackets are excited along the lower adiabatic LIP to the higher quasibound states, and the other is  that the inherent nonadibatic coupling $V_{{\nu JM}E}$ in Eq. (\ref{PE1}) may suppress the lower quasibound states to cross over the avoided crossing region.\\ \indent
 \begin{figure*}[!t]\centering
\resizebox{0.8\textwidth}{!}{%
  \includegraphics{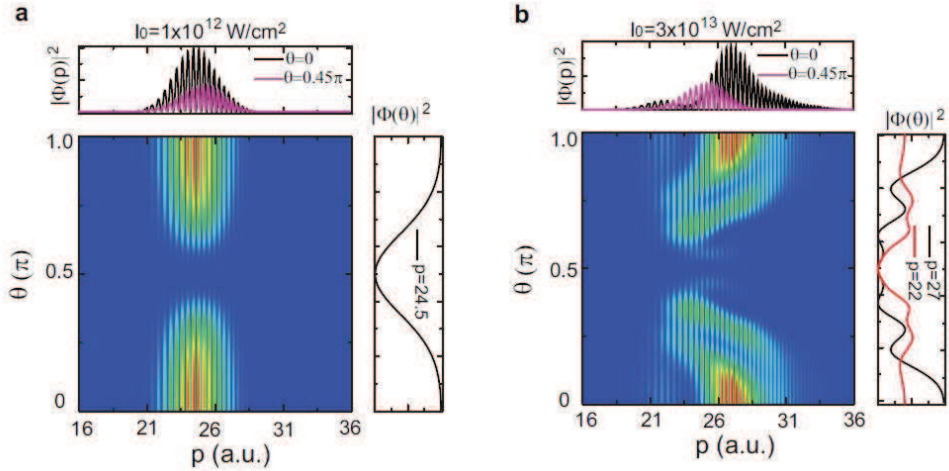}
} \caption{Probability distributions for relative momentum of Na+I. The results, i.e., $|\Phi(p, \theta, t)|^2$, are observed at  $t=1.2$ ps, corresponding to two outgoing fragments with  \textbf{a} the weak-field case at $I_0=1.0\times10^{12}$ W/cm$^2$, and \textbf{b}  the strong field case at $I_0=3.0\times10^{13}$ W/cm$^2$.}\label{fig4}
\end{figure*}
\begin{figure}[!t]\centering
\resizebox{0.8\textwidth}{!}{%
  \includegraphics{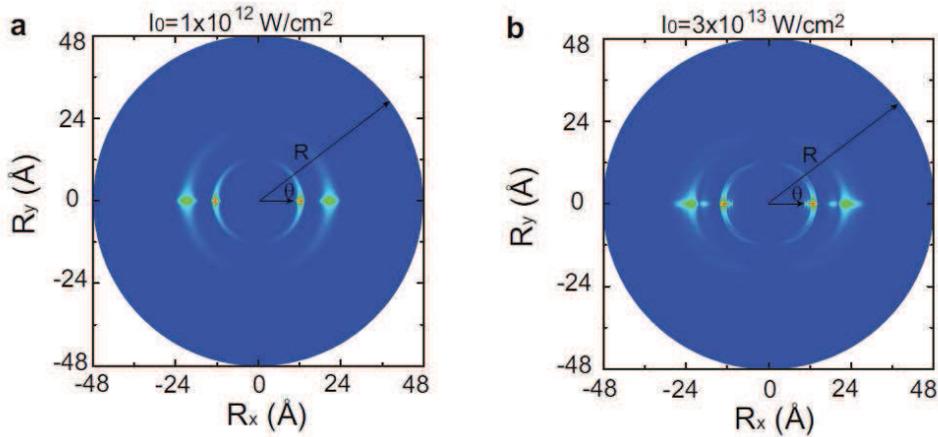}
} \caption{Probability distributions for the wavepackets of photofragments. The results are observed in the asymptotic dissociation region $|g(R)\Psi_e(R, \theta, t)|^2$ at $t=1.2$ ps, with  \textbf{a} the weak-field case at $I_0=1.0\times10^{12}$ W/cm$^2$, and \textbf{b}  the strong field case at $I_0=3.0\times10^{13}$ W/cm$^2$.}\label{fig5}
\end{figure}
We now examine the fragment distributions at $t=1.2$ ps in Figures \ref{fig4} and $\ref{fig5}$, corresponding to  two outgoing fragments.  As indicated in  Eq. (\ref{PE1}), quantum interferences  in the momentum distributions are found in both the weak and the strong field cases. The spacing between the momentum lines around the center corresponds to
$\sim59$ cm$^{?1}$,  in good agreement with the known energy spacing between the quasibound energy levels of NaI$^\ast$. Since the transient momentum spectrum is calculated  after  the laser pulse,  the observed energy spacing is independent of laser intensity. As shown in Figure \ref{fig5}, two well-separated wave-packets appear in the dissociation asymptotic region. Since the two space-separated wavepackets come from the same  ``source", i.e., the reaction intermediate NaI$^{\ast}$ associated with the trapped wavepacket, they can create quantum interference in the momentum space.
 In the region near $\theta=\pm\pi/2$, the separated wavepackets only show a single-``crescent-like" distribution, and the corresponding momentum distributions are again similar to the weak field case.  The interference patterns observed in Figure \ref{fig4} provide a direct signature to distinguish  the indirect dissociation reaction from the direct one, where the excited molecule will not be trapped under the field-free condition. Note that the wavepacket after 1.2 ps (as shown in Figure \ref{fig5}) is just spread to the region of $R\leq36$ {\AA} far from the boundary, indicating that there is no influence  of the absorbing potential on the accuracy of the calculations.  \\ \indent
 It is interesting to note that in the strong field case the angular distributions of the photofragments at a given momentum become complex, going beyond those in the weak field case.
 This rotational structure can be explained by Eq. (\ref{PE1}), i.e., due to the strong-field-molecule interaction,  more than one rotational state associated with higher order $P_J^M(\cos\theta)$ contribute to the fragment distributions.  This mechanism is different from quantum interference effects caused by the existence of the LICI, which has been observed in direct photofragmentation reactions for example in the molecule ions, H$_2^+$ and D$_2^+$ \cite{jpcl:6:348,prl:116:143004}. In the present case for the molecule NaI,  the field-molecule interaction time ($<100$ fs), as compared with the rotational period of the excited wavepackets (basic rotational period of NaI is 138 ps),  is too short to rotate the molecule in the upper LIP to the vicinity of the LICI.  As a result, there is almost no the excited wavepacket that undergos nonadiabatic transitions from the upper LIP to the higher quasibound states which can interfere with the wavepacket along the  lower LIP $V_g^{LIP}$.\\ \indent
 In summary, we have  investigated  how strong-laser-field-modified molecular potentials play a role in indirect photofragmentation reactions. A theoretical analysis combined with the time-dependent wavepacket simulations was applied to the photochemical  reactions of  breaking up the molecule NaI into  the neutral atoms Na and I through the reaction intermediate NaI$^{\ast}$. We have identified that  the photofragment distributions  are produced upon laser-induced quasibound states NaI$^{\ast}$, and  shown that the energies of these quasibound states  NaI$^{\ast}$ can be extracted from the transient photofragment momentum distributions. These results provide a new approach to gain an insight into the reaction intermediate directly from photochemical reaction products, and this method in principle can be extended to study more complex molecular systems. Although the present work highlighted how a single strong laser field affects the reaction intermediate,  it is also interesting to explore how commonly used coherent control schemes can be employed to control the reaction intermediate. To this end, we may use a second either resonant or nonresonant pulse to further excite the wavepacket of the reaction intermediate. Furthermore, the scheme presented in this work can also be transposed to examine indirect dissociation in the case of photo-predissociation involving multiple electronic excited states.
\section*{Author information}
\subsection*{Corresponding Author}
$\ast$(C.C.S) Email:c.shu@unsw.edu.au\\
$\ast$(A.D.B) Email:andre.bandrauk@usherbrooke.ca
\subsection*{Notes}
The authors declare no competing financial interest.
\begin{acknowledgement}
C.C.S. acknowledges the financial support by the Vice-Chancellor's Postdoctoral Research Fellowship of The University of New South Wales, Australia. K.J.Y.  acknowledges the support and hospitality provided by UNSW  Canberra during his visit.  D.D. and I.R.P. acknowledge partial supports  by the Australian Research Council under Grant Nos. DP130101658 and FL110100020.
The authors also thank RQCHP and Compute Canada for access to massively parallel computer clusters.
\end{acknowledgement}

\end{document}